\documentclass[12pt]{article}
\begin{document}
\setcounter{page}{1}
\def\theequation{\arabic{section}.\arabic{equation}}
\def\theequation{\thesection.\arabic{equation}}
\setcounter{section}{0}

\title{On the polarization properties of the charmed baryon
$\Lambda^+_c$ in the $\Lambda^+_c \to p + K^- + \pi^+ + \pi^0$ decay}

\author{A. Ya. Berdnikov, Ya. A. Berdnikov\,\thanks{E--mail:
berdnikov@twonet.stu.neva.ru} , A. N. Ivanov, \\ V. F. Kosmach,\\
M. D. Scadron\,\thanks{E--mail: scadron@physics.arizona.edu, Physics
Department, University of Arizona, Tucson, Arizona 85721, USA} , and
N. I. Troitskaya}

\date{\today}

\maketitle

\begin{center}
{\it State Technical University of St. Petersburg, Department of
Nuclear Physics, \\ Polytechnicheskaya 29, 195251 St. Petersburg, Russian
Federation}
\end{center}

\begin{center}
\begin{abstract}
The polarization properties of the charmed $\Lambda^+_c$ baryon are
investigated in weak non--leptonic four--body $\Lambda^+_c \to p + K^-
+ \pi^+ + \pi^0$ decay. The probability of this decay and the angular
distribution of the probability are calculated in the effective quark
model with chiral $U(3)\times U(3)$ symmetry incorporating Heavy Quark
Effective theory (HQET) and the extended Nambu--Jona--Lasinio model
with a linear realization of chiral $U(3)\times U(3)$ symmetry. The
theoretical value of the probability of the decay $\Lambda^+_c \to p +
K^- + \pi^+ + \pi^0$ relative to the probability of the decay
$\Lambda^+_c \to p + K^- + \pi^+$ does not contain free parameters and
fits well experimental data. The application of the obtained results
to the analysis of the polarization of the $\Lambda^+_c$ produced in
the processes of photo and hadroproduction is discussed.
\end{abstract}
\end{center}

\newpage

\section{Introduction}
\setcounter{equation}{0}

It is known that in reactions of photo and
hadroproduction the charmed baryon $\Lambda^+_c$ is produced polarized
[1]. The analysis of the $\Lambda^+_c$ polarization via the
investigation of the decay products should give an understanding of
the mechanism of the charmed baryon production at high energies.

Recently [2] we have given a theoretical analysis of the polarization
properties of the $\Lambda^+_c$ in the mode $\Lambda^+_c \to p + K^- +
\pi^+$. This is the most favourable mode of the $\Lambda^+_c$ decays
from the experimental point of view. From the theoretical point of
view this mode is the most difficult case of the analysis of the weak
non--leptonic decays of the $\Lambda^+_c$ baryon [1,2]. Indeed, for the
calculation of the matrix element of the transition $\Lambda^+_c \to p
+ K^- + \pi^+$ the baryonic and mesonic degrees of freedom cannot be
fully factorized.

In spite of these theoretical difficulties the problem of the
theoretical analysis of the decay $\Lambda^+_c \to p + K^- + \pi^+$
has been successfully solved within the effective quark model with
chiral $U(3)\times U(3)$ symmetry incorporating Heavy Quark Effective
Theory (HQET) [3,4] and the extended Nambu--Jona--Lasinio (ENJL) model
with a linear realization of chiral $U(3)\times U(3)$ symmetry
[5--7]\,\footnote{All results obtained below are valid for the Linear
Sigma Model (L$\sigma$M) [8] supplemented by HQET as well.}. Such an
effective quark model with chiral $U(3)\times U(3)$ symmetry motivated
by the low--energy effective QCD with a linearly rising interquark
potential responsible for a quark confinement [9] describes well
low--energy properties of light and heavy mesons [5,6] as well as the
octet and decuplet of light baryons [7].

In the effective quark model with chiral $U(3)\times U(3)$ symmetry
(i) baryons are the three--quark states [10] and do not contain any
bound diquark states, then (ii) the spinorial structure of the
three--quark currents is defined as the products of the axial--vector
diquark densities $[\bar{q^c}_i(x)\gamma^{\mu}q_j(x)]$ and a quark
field $q_k(x)$ transforming under $SU(3)_f\times SU(3)_c$ group like
$(\underline{6}_f,\tilde{\underline{3}}_c)$ and
$(\underline{3}_f,\underline{3}_c)$ multiplets, respectively, where
$i,j$ and $k$ are the colour indices running through $i=1,2,3$ and $q
= u,d$ or $s$ quark field. This agrees with the structure of the
three--quark currents used for the investigation of the properties of
baryons within QCD sum rules approach [11]. As has been shown in
Ref.[9] this is caused by the dynamics of strong low--energy
interactions imposed by a linearly rising interquark potential. The
fixed structure of the three--quark currents allows to describe all
variety of low--energy interactions of baryon octet and decuplet in
terms of the phenomenological coupling constant $g_{\rm B}$.  The
coupling constants $g_{\rm \pi NN}$, $g_{\rm \pi N \Delta}$ and
$g_{\rm \gamma N \Delta}$ interactions, and the $\sigma_{\rm \pi
N}$--term of the low--energy ${\rm \pi N}$--scattering have been
calculated in good agreement with the experimental data and other
phenomenological approaches based on QCD [7,12].

In this paper we apply the effective quark model with chiral
$U(3)\times U(3)$ symmetry [2,5--7] to the investigation of the
polarization properties of of the $\Lambda^+_c$ baryon in weak
non--leptonic four--body decays and treat the most favourable
experimentally four--body mode $\Lambda^+_c \to p + K^- + \pi^+ +
\pi^0$. The experimental value of the probability of this decay is
equal to [13]
\begin{eqnarray}\label{label1.1}
B(\Lambda^+_c \to p K^- \pi^+ \pi^0)_{\exp} = (3.4\pm
1.0)\,\%.
\end{eqnarray}
Relative to the decay $\Lambda^+_c$ $\to$ p + K$^-$ + $\pi^+$ the
experimental probability of which is $B(\Lambda^+_c \to p K^-
\pi^+) = 0.050\pm 0.013$ [13] the probability of the decay 
$\Lambda^+_c \to p + K^- + \pi^+ + \pi^0$ reads
\begin{eqnarray}\label{label1.2}
B(\Lambda^+_c \to p K^- \pi^+ \pi^0/\Lambda^+_c \to p K^-
\pi^+)_{\exp} = (0.68\pm 0.27).
\end{eqnarray}
We would like to emphasize that the weak non--leptonic four--body mode
$\Lambda^+_c \to p + K^- + \pi^+ + \pi^0$ as well as the mode
$\Lambda^+_c \to p + K^- + \pi^+$ is rather difficult for the
theoretical analysis [1,2], since baryonic and mesonic degrees of
freedom cannot be fully factorized.

For the theoretical analysis of the weak non--leptonic decays of the
$\Lambda^+_c$ baryon we would use the effective low--energy Lagrangian
[2] (see also Refs.[12,14])
\begin{eqnarray}\label{label1.3}
{\cal L}_{\rm eff}(x) &=& -\frac{G_F}{\sqrt{2}}\,V^*_{c s}\,V_{u
d}\,\Big\{C_1(\Lambda_{\chi})\,[\bar{s}(x)\,
\gamma^{\mu}(1-\gamma^5)\,c(x)]\,[\bar{u}(x)\,
\gamma_{\mu}(1-\gamma^5)\,d(x)]\nonumber\\
&& \hspace{0.7in} +
C_2(\Lambda_{\chi})\,[\bar{u}(x)\,
\gamma^{\mu}(1-\gamma^5)\,c(x)]\,[\bar{s}(x)\,
\gamma_{\mu}(1-\gamma^5)\,d(x)]\Big\},
\end{eqnarray}
where $G_F=1.166\times 10^{-5}\;{\rm GeV}^{-2}$ is the Fermi weak
constant, $V^*_{c s}$ and $V_{u d}$ are the elements of the
CKM--mixing matrix, $C_i(\Lambda_{\chi})\,(i=1,2)$ are the Wilson
coefficients caused by the strong quark--gluon interactions at scales
$p > \Lambda_{\chi}$ (short--distance contributions), where
$\Lambda_{\chi} = 940\,{\rm MeV}$ is the scale of spontaneous breaking
of chiral symmetry (SB$\chi$S) [2,5--7]. The numerical values of the
coefficients $C_1(\Lambda_{\chi}) = 1.24$ and $C_2(\Lambda_{\chi}) =
-0.47$ have been calculated in Ref.[2].

Following Ref.[2] for the calculation of the probability of the decay
$\Lambda^+_c \to p + K^- + \pi^+ + \pi^0$ we suggest to use the
effective Lagrangian Eq.(\ref{label1.3}) reduced to the form
\begin{eqnarray}\label{label1.4}
{\cal L}_{\rm eff}(x) = -\frac{G_F}{\sqrt{2}}\,V^*_{c s}\,V_{u
d}\,\bar{C}_1(\Lambda_{\chi})\,[\bar{s}(x)\,\gamma_{\mu}
(1-\gamma^5)\,c(x)]\,[\bar{u}(x)\,\gamma^{\mu}(1-\gamma^5)\,d(x)]
\end{eqnarray}
by means of a Fierz transformation [2], where
$\bar{C}_1(\Lambda_{\chi}) = C_1(\Lambda_{\chi}) + C_2
(\Lambda_{\chi})/N$ with $N=3$, the number of quark colour degrees of
freedom\footnote{We would like to accentuate that our approach to
non--leptonic decays of charmed baryons agrees in principle with the
current--algebra analysis of non--leptonic decays of light and charmed
baryons based on $(V-A)\times (V-A)$ effective coupling developed by
Scadron {\it et al.} in Refs.[15].}.

The paper is organized as follows. In Sect.\,2 we calculate the
amplitude of the decay mode $\Lambda^+_c \to p + K^- + \pi^+ + \pi^0$.
In Sect.\,3 we calculate the angular distribution of the probability
and the probability of the decay $\Lambda^+_c \to p + K^- + \pi^+ +
\pi^0$ relative to the probability of the $\Lambda^+_c \to p + K^- +
\pi^+$. In Sect.\,4 we analyse the polarization properties of the
charmed baryon $\Lambda^+_c$.  In the Conclusion we discuss the
obtained results.

\section{Amplitude of the $\Lambda^+_c \to
p + K^- + \pi^+ + \pi^0$ decay} 
\setcounter{equation}{0}

The amplitude of the decay $\Lambda^+_c \to p + K^- +
\pi^+ + \pi^0$ decay we define in the usual way [2,12]
\begin{eqnarray}\label{label2.1}
\frac{\displaystyle{\cal M}(\Lambda^+_c(Q) \to p(q)
K^-(q_1)\pi^+(q_2)\pi^0(q_3))}{\displaystyle
\sqrt{2E_{\Lambda^+_c}V\,2E_p V\,2E_{K^-} V\,2E_{\pi^+} V\,2E_{\pi^0}
V}} = \langle p(q) K^-(q_1)\pi^+(q_2)\pi^0(q_3)|{\cal L}_{\rm
eff}(0)|\Lambda^+_c(Q)\rangle ,
\end{eqnarray}
where $E_i\,(i=\Lambda^+_c,p,K^-,\pi^+,\pi^0)$ are the energies of the
$\Lambda^+_c$, the proton and mesons, respectively.

Since experimentally the probability of the decay mode $\Lambda^+_c
\to p + K^- + \pi^+ + \pi^0$ is measured relative to the probability
of the $\Lambda^+_c \to p + K^- + \pi^+$ decay, so that we would treat
it with respect to the probability of the decay $\Lambda^+_c \to p +
K^- + \pi^+$ the partial width of which has been calculated in Ref.[2]
and reads 
\begin{eqnarray}\label{label2.2}
\hspace{-0.3in}\Gamma(\Lambda^+_c \to p\,K^- \pi^+) = |G_F\,V^*_{c
s}\,V_{u d}\,\bar{C}_1(\Lambda_{\chi})|^2 \,\Bigg[g_{\rm \pi
NN}\,\frac{4}{5}\,\frac{g_{\rm C}}{g_{\rm
B}}\,\frac{F_{\pi}\Lambda_{\chi}}{m^2}\Bigg]^2\,\times\,\Bigg[\frac{5
M^5_{\Lambda^+_c}}{512\pi^3}\Bigg]\,\times\,f(\xi).
\end{eqnarray}
The function $f(\xi)$ is determined by the integral [2]
\begin{eqnarray}\label{label2.3}
\hspace{-0.5in}f(\xi) = \int\limits^{1 + \xi^2/4}_{\xi}\Bigg(1 -
\frac{3}{5}\,x + \frac{2}{15}\,x^2 + \frac{7}{60}\,\xi^2 -
\frac{2}{5}\,\frac{\xi^2}{x}\Bigg)\,x\,\sqrt{x^2 - \xi^2}\,dx = 0.065,
\end{eqnarray}
where $\xi = 2M_p/M_{\Lambda^+_c}$. The numerical value has been
obtained at $M_{\Lambda^+_c}=2285\,{\rm MeV}$ and $M_p = 938\,{\rm
MeV}$, the mass of the $\Lambda^+_c$ baryon and the proton,
respectively, and in the chiral limit, i.e. at zero masses of daugther
mesons.  The coupling constants $g_{\rm B}$ and $g_{\rm C}$ determine
the interactions of the proton and the $\Lambda^+_c$ baryon with the
three--quark currents $\eta_{\rm N}(x) =
-\varepsilon^{ijk}[\bar{u^c}_i(x)\gamma^{\mu}u_j(x)]\gamma_{\mu}\gamma^5
d_k(x)$ and $\bar{\eta}_{\Lambda^+_c}(x) =
\varepsilon^{ijk}\bar{c}_i(x)\gamma_{\mu}\gamma^5[\bar{d}_j(x)
\gamma^{\mu}u^c_k(x)]$, respectively [2,7]:
\begin{eqnarray}\label{label2.4}
\hspace{-0.5in}{\cal L}_{\rm int}(x) = \frac{g_{\rm
B}}{\sqrt{2}}\,\bar{\psi}_p(x)\,\eta_{\rm N}(x) + \frac{g_{\rm
C}}{\sqrt{2}}\,\bar{\eta}_{\Lambda^+_c}(x)\,\psi_{\Lambda^+_c}(x) +
{\rm h.c.}.
\end{eqnarray}
Here $\psi_p(x)$ and $\psi_{\Lambda^+_c}(x)$ are the interpolating
fields of the proton and the $\Lambda^+_c$ baryon. The coupling
constant $g_{\rm B}$ has been related in Ref.[7] to the quark
condensate $\langle \bar{q}(0)q(0)\rangle = -\,(255\,{\rm MeV})^3$,
the constituent quark mass $m = 330\,{\rm MeV}$ calculated in the
chiral limit\footnote{This agrees with the results obtained by Elias
and Scadron [16].}, the leptonic coupling constant $F_{\pi} =
92.4\,{\rm MeV}$ of pions calculated in the chiral limit, the ${\rm
\pi NN}$ coupling constant $g_{\rm \pi NN} = 13.4$ and as well as the
mass of the proton $M_{\rm p}$:
\begin{eqnarray}\label{label2.5}
g_{\rm \pi NN} = g^2_{\rm B}\,\frac{2m}{3F_{\pi}}\,\frac{\langle
\bar{q}(0)q(0)\rangle^2}{M^2_p}.
\end{eqnarray}
Numerically $g_{\rm B}$ is equal to $g_{\rm B} =1.34\times
10^{-4}\,{\rm MeV}$ [7]. The coupling constant $g_{\rm C}$ has been
fixed in Ref.[2] through the experimental value of the partial width
of the decay $\Lambda^+_c \to p + K^- + \pi^+$. The coupling constant
$g_{\rm C}$ appears in all partial widths of the decay
modes of the $\Lambda^+_c$ baryon and cancels itself in the ratio
\begin{eqnarray}\label{label2.6}
B(\Lambda^+_c \to pK^-\pi^+\pi^0/\Lambda^+_c \to pK^-\pi^+) =
\frac{\Gamma(\Lambda^+_c \to p K^-\pi^+\pi^0)}{\Gamma(\Lambda^+_c \to
p\,K^- \pi^+)}.
\end{eqnarray}
The amplitude of the decay $\Lambda^+_c \to p + K^- + \pi^+ + \pi^0$
we calculate in the tree--meson approximation and in the chiral limit
[2]
\begin{eqnarray}\label{label2.7}
\hspace{-0.5in}&&\frac{\displaystyle{\cal M}(\Lambda^+_c(Q) \to p(q)
K^-(q_1)\pi^+(q_2)\pi^0(q_3))}{\displaystyle
\sqrt{2E_{\Lambda^+_c}V\,2E_p V\,2E_{K^-}
V\,2E_{\pi^+}V\,2E_{\pi^0}V}} = \langle p(q)
K^-(q_1)\pi^+(q_2)\pi^0(q_3)|{\cal L}_{\rm
eff}(0)|\Lambda^+_c(Q)\rangle =\nonumber\\
\hspace{-0.5in}&&= -\frac{G_F}{\sqrt{2}}\,
V^*_{c s}\,V_{u d}\,\bar{C}_1(\Lambda_{\chi})\,
\langle p(q) K^-(q_1)|\bar{s}(0) 
\gamma_{\mu}(1-\gamma^5) c(0)|\Lambda^+_c(Q)\rangle \nonumber\\
\hspace{-0.5in}&&\times \langle
\pi^+(q_2)\pi^0(q_3)|\bar{u}(0)\,\gamma^{\mu}(1-\gamma^5)\,d(0)|0\rangle
,
\end{eqnarray}
The matrix element of the transition $\Lambda^+_c \to p + K^-$ has 
been calculated in Ref.[2] and reads
\begin{eqnarray}\label{label2.8}
\hspace{-0.5in}&&\sqrt{2E_{\Lambda^+_c}V\,2E_p V\,2E_{K^-} V}\,\langle
p(q)
K^-(q_-)|\bar{s}(0)\,\gamma_{\mu}(1-\gamma^5)\,c(0)|\Lambda^+_c(Q)\rangle
= \nonumber\\
\hspace{-0.5in}&&= i g_{\rm \pi NN}\,\frac{4}{5}\,\frac{g_{\rm
C}}{g_{\rm
B}}\,\frac{\Lambda_{\chi}}{m^2}\,\bar{u}_p(q,\sigma^{\prime}\,)
\,[2\,v_{\mu}(1 - \gamma^5) + \gamma_{\mu}(1 +
\gamma^5)]\,u_{\Lambda^+_c}(Q,\sigma) =\nonumber\\
\hspace{-0.5in}&&= i g_{\rm \pi NN}\,\frac{4}{5}\,
\frac{g_{\rm C}}{g_{\rm B}}\,\frac{\Lambda_{\chi}}{m^2}\,
\bar{u}_p(q,\sigma^{\prime}\,) \,(1 - \gamma^5)\,
(2\,v_{\mu} + \gamma_{\mu})\,u_{\Lambda^+_c}(Q,\sigma),
\end{eqnarray}
where $\bar{u}_p(q,\sigma^{\prime}\,)$ and $u_{\Lambda^+_c}(Q,\sigma)$
are the Dirac bispinors of the proton and the $\Lambda^+_c$ baryon,
$v^{\mu}$ is a 4--velocity of the $\Lambda^+_c$ baryon defined by
$Q^{\mu} = M_{\Lambda^+_c}\,v^{\mu}$.

The matrix element of the transition $0 \to \pi^+ + \pi^0$ has been
calculated in [5] and reads
\begin{eqnarray}\label{label2.9}
\sqrt{2E_{\pi^+}V\,2E_{\pi^0}V} \langle\pi^+(q_2)\pi^0(q_3)|
\bar{u}(0)\gamma^{\mu}(1-\gamma^5)\,d(0)|0\rangle = - \sqrt{2}\,(q_2 -
q_3)^{\mu}.
\end{eqnarray}
Hence, the amplitude of the decay $\Lambda^+_c \to p + K^- + \pi^+ +
\pi^0$ is given by
\begin{eqnarray}\label{label2.10}
\hspace{-0.7in}&&{\cal M}(\Lambda^+_c(Q) \to p(q)
K^-(q_1)\pi^+(q_2)\pi^0(q_3)) = i\,G_F\,V^*_{c s}\,V_{u
d}\,\bar{C}_1(\Lambda_{\chi})\,\nonumber\\
\hspace{-0.7in}&&\times \,\frac{4}{5}\,\frac{g_{\rm \pi
NN}}{M_{\Lambda^+_c}}\,\Bigg[\frac{g_{\rm C}}{g_{\rm
B}}\,\frac{\Lambda_{\chi}}{m^2}\Bigg]\,\bar{u}_p(q,\sigma^{\prime}\,)
\,(1 - \gamma^5)\,[2 Q\cdot (q_2 - q_3) + M_{\Lambda^+_c}(\hat{q}_2 -
\hat{q}_3)]\,u_{\Lambda^+_c}(Q,\sigma).
\end{eqnarray}
Now we can proceed to the evaluation of the probability of the
$\Lambda^+_c \to p + K^- + \pi^+ + \pi^0$ decay.

\section{Probability and angular distribution of the decay 
$\Lambda^+_c \to p + K^- + \pi^+ + \pi^0$} 
\setcounter{equation}{0}

The differential partial width of the $\Lambda^+_c \to
p + K^- + \pi^+ + \pi^0$ decay is determined by
\begin{eqnarray}\label{label3.1}
\hspace{-0.5in}&&d\Gamma(\Lambda^+_c \to p K^- \pi^+ \pi^0) =
\frac{1}{2M_{\Lambda^+_c}}\,\overline{|{\cal M}(\Lambda^+_c(Q) \to p(q)
K^-(q_1)\pi^+(q_2)\pi^0(q_3))|^2}\nonumber\\
\hspace{-0.5in}&&\times\,(2\pi)^4\,\delta^{(4)}(Q - q - q_1 - q_2 -
q_3)\,\frac{d^3q}{(2\pi)^3 2 E_p}\,\frac{d^3q_1}{(2\pi)^3 2
E_{K^-}}\,\frac{d^3q_2}{(2\pi)^3 2 E_{\pi^+}}\,\frac{d^3q_1}{(2\pi)^3
2 E_{\pi^0}}.
\end{eqnarray}
We calculate the quantity $\overline{|{\cal M}(\Lambda^+_c(Q) \to p(q)
K^-(q_1)\pi^+(q_2)\pi^0(q_3))|^2}$ for the polarized $\Lambda^+_c$ and
unpolarized proton
\begin{eqnarray}\label{label3.2}
\hspace{-0.5in}&&\overline{|{\cal M}(\Lambda^+_c(Q) \to p(q)
K^-(q_1)\pi^+(q_2)\pi^0(q_3))|^2} = |G_F\,V^*_{c s}\,V_{u
d}\,\bar{C}_1(\Lambda_{\chi})|^2\,\left[\frac{4}{5}\,\frac{g_{\rm \pi
NN}}{M_{\Lambda^+_c}}\,\frac{g_{\rm C}}{g_{\rm
B}}\,\frac{\Lambda_{\chi}}{m^2}\right]^2\nonumber\\
\hspace{-0.5in}&&\times\,\frac{1}{2}\,{\rm tr}\{(M_{\Lambda^+_c} +
\hat{Q})(1 + \gamma^5\hat{\omega}_{\Lambda^+_c})[2Q\cdot(q_2-q_3) +
M_{\Lambda^+_c}(\hat{q}_2 - \hat{q}_3)](1+\gamma^5)(M_p +
\hat{q})(1-\gamma^5)\nonumber\\
\hspace{-0.5in}&&\times\,[2Q\cdot(q_2-q_3) + M_{\Lambda^+_c}(\hat{q}_2
- \hat{q}_3)]\},
\end{eqnarray}
where $\omega^{\mu}_{\Lambda^+_c}$ is a space--like unit vector,
$\omega^2_{\Lambda^+_c} = - 1$, orthogonal to the 4--momentum of the
$\Lambda^+_c$, $Q\cdot \omega_{\Lambda^+_c} = 0$. It is related to the
direction of the $\Lambda^+_c$ spin defined by 
\begin{eqnarray}\label{label3.3}
\omega^{\mu}_{\Lambda^+_c} =\left(\frac{\displaystyle \vec{Q}\cdot
\vec{\omega}_{\Lambda^+_c}}{\displaystyle M_{\Lambda^+_c}},
\vec{\omega}_{\Lambda^+_c} + \frac{\displaystyle \vec{Q}(\vec{Q}\cdot
\vec{\omega}_{\Lambda^+_c})}{\displaystyle
M_{\Lambda^+_c}(E_{\Lambda^+_c} + M_{\Lambda^+_c})}\right),
\end{eqnarray}
where $\vec{\omega}^{\,2}_{\Lambda^+_c} = 1$. At the rest frame of the
$\Lambda^+_c$ we have $\omega^{\mu}_{\Lambda^+_c} = (0,
\vec{\omega}_{\Lambda^+_c})$.

For the differential branching ratio $B(\Lambda^+_c \to p K^- \pi^+
\pi^0/\Lambda^+_c \to pK^-\pi^+)$ defined by Eq.(\ref{label2.6}) we
get
\begin{eqnarray}\label{label3.4}
\hspace{-0.5in}&&dB(\Lambda^+_c \to p K^- \pi^+ \pi^0/\Lambda^+_c \to
pK^-\pi^+) = \frac{1024\pi^3}{1.3 M^8_{\Lambda^+_c}}\,
\frac{1}{F^2_{\pi}}\,\frac{1}{2}\,{\rm
tr}\{(M_{\Lambda^+_c} + \hat{Q})(1 +
\gamma^5\hat{\omega}_{\Lambda^+_c})\nonumber\\
\hspace{-0.5in}&&\times\,[2Q\cdot(q_2-q_3) + M_{\Lambda^+_c}(\hat{q}_2
- \hat{q}_3)](1+\gamma^5)(M_p + \hat{q})(1-\gamma^5)\nonumber\\
\hspace{-0.5in}&&\times\,[2Q\cdot(q_2-q_3) + M_{\Lambda^+_c}(\hat{q}_2
- \hat{q}_3)]\}\,(2\pi)^4\,\delta^{(4)}(Q - q - q_1 - q_2 -
q_3)\nonumber\\
\hspace{-0.5in}&&\times\,\frac{d^3q}{(2\pi)^3 2 E_p}\,\frac{d^3q_1}{(2\pi)^3 2
E_{K^-}}\,\frac{d^3q_2}{(2\pi)^3 2 E_{\pi^+}}\,\frac{d^3q_3}{(2\pi)^3
2 E_{\pi^0}}.
\end{eqnarray}
The trace amounts to
\begin{eqnarray}\label{label3.5}
\frac{1}{2}\,{\rm tr}\{\ldots\}&=& 16\,Q\cdot q\,(Q\cdot
(q_2-q_3))^2\nonumber\\ &+& M_{\Lambda^+_c}[16\,Q\cdot q\,Q\cdot
(q_2-q_3)\,(q_2-q_3)\cdot \omega_{\Lambda^+_c} - 32\,(Q\cdot
(q_2-q_3))^2\,q\cdot \omega_{\Lambda^+_c}]\nonumber\\ &+&
M^2_{\Lambda^+_c}[24\,Q\cdot (q_2-q_3)\,q\cdot (q_2-q_3) -
4\,Q\cdot q\,(q_2 - q_3)^2]\nonumber\\ &+&
M^3_{\Lambda^+_c}[8\,q\cdot (q_2-q_3)\,(q_2-q_3)\cdot
\omega_{\Lambda^+_c} - 4\,(q_2 - q_3)^2\,q\cdot
\omega_{\Lambda^+_c}].
\end{eqnarray}
For the integration over the momenta of $\pi$ mesons it is useful to
apply the formula [2]
\begin{eqnarray}\label{label3.6}
\hspace{-0.5in}\int (q_2 - q_3)_{\alpha}(q_2 -
q_3)_{\beta}\,\delta^{(4)}(P -q_2 - q_3)
\frac{d^3q_2}{2E_{\pi^+}}\frac{d^3q_3}{2E_{\pi^0}}= \frac{\pi}{6}\,
\Big(- P^2\,g_{\alpha\beta} + P_{\alpha}P_{\beta}\Big),
\end{eqnarray}
where $P = Q - q - q_1$. Integrating over the momenta of pions we
arrive at the following expression for the differential branching
ratio $B(\Lambda^+_c \to p K^- \pi^+ \pi^0/\Lambda^+_c \to
pK^-\pi^+)$:
\begin{eqnarray}\label{label3.7}
\hspace{-0.5in}&&dB(\Lambda^+_c \to p K^- \pi^+ \pi^0/\Lambda^+_c \to
pK^-\pi^+) = \frac{2.1}{4\pi^4}\,\frac{1}{M^8_{\Lambda^+_c}}\,
\frac{1}{F^2_{\pi}}\,\{4\,Q\cdot q\,((Q\cdot P)^2 - Q^2P^2)\nonumber\\
\hspace{-0.5in}&& + M_{\Lambda^+_c}\,
[4\,Q\cdot q\, Q\cdot
P\, P\cdot \omega_{\Lambda^+_c} - 8\,((Q\cdot P)^2 - Q^2P^2)\,
 P\cdot \omega_{\Lambda^+_c}]+ M^2_{\Lambda^+_c}\,
(- 3\,Q\cdot q\,P^2\nonumber\\
\hspace{-0.5in}&& + 6\,Q\cdot P\,q\cdot P) +
M^3_{\Lambda^+_c}\,(P^2\,q\cdot \omega_{\Lambda^+_c} + 2\,q\cdot
P\,P\cdot \omega_{\Lambda^+_c})\}\,\frac{d^3q}{E_p}\,\frac{d^3q_1}{
E_{K^-}}.
\end{eqnarray}
After the integration over the momenta of the K$^-$ meson and the
energies of the proton we obtain the angular distribution of the
probability of the decay mode $ \Lambda^+_c \to p + K^- + \pi^+ +
\pi^0$ relative to the probability of the decay $\Lambda^+_c \to
p + K^- + \pi^+$ in the rest frame of the $\Lambda^+_c$ baryon:
\begin{eqnarray}\label{label3.8}
\hspace{-0.5in}4\pi\,\frac{dB}{d\Omega_{\vec{n}_p}}(\Lambda^+_c \to p
K^- \pi^+ \pi^0/\Lambda^+_c \to pK^-\pi^+) = 0.87\,(1 -
0.09\,\vec{n}_p\cdot\vec{\omega}_{\Lambda^+_c}),
\end{eqnarray}
where $\vec{n}_p = \vec{q}/|\vec{q}\,|$ is a unit vector directed
along the momentum of the proton and $\Omega_{\vec{n}_p}$ is the solid
angle of the unit vector $\vec{n}_p$.

Integrating the angular distribution Eq.(\ref{label3.8}) over the
solid angle $\Omega_{\vec{n}_p}$ we obtain the total branching ratio
\begin{eqnarray}\label{label3.9}
B(\Lambda^+_c \to p K^- \pi^+ \pi^0/\Lambda^+_c \to pK^-\pi^+) = 0.87.
\end{eqnarray}
The theoretical value fits well the experimental data
Eq.(\ref{label1.2}): $B(\Lambda^+_c \to p K^- \pi^+ \pi^0/\Lambda^+_c
\to p K^- \pi^+)_{\exp} = (0.68\pm 0.27)$.

\section{Polarization of the charmed baryon $\Lambda^+_c$}
\setcounter{equation}{0}

The formula Eq.(\ref{label3.8}) describes the polarization of the
charmed $\Lambda^+_c$ baryon relative to the momentum of the proton in
the decay mode $\Lambda^+_c \to p + K^- + \pi^+ + \pi^0$.  If the spin
of the $\Lambda^+_c$ is parallel to the momentum of the proton, the
right--handed (R) polarization, the scalar product
$\vec{\omega}_{\Lambda^+_c}\cdot \vec{n}_p$ amounts to
$\vec{\omega}_{\Lambda^+_c}\cdot \vec{n}_p = \cos\vartheta$. The
angular distribution of the probability reads
\begin{eqnarray}\label{label4.1}
4\pi\,\frac{dB}{d\Omega_{\vec{n}_p}}(\Lambda^+_c \to p K^- \pi^+
\pi^0/\Lambda^+_c \to pK^-\pi^+)_{\rm (R)} = 0.87\,(1 -
0.09\,\cos\vartheta).
\end{eqnarray}
In turn, for the left--handed (L) polarization of the $\Lambda^+_c$,
the spin of the $\Lambda^+_c$ is anti--parallel to the momentum of the
proton, the scalar product reads $(\vec{\omega}_{\Lambda^+_c}\cdot
\vec{n}_p) = -\cos\vartheta$ and the angular distribution becomes
equal to
\begin{eqnarray}\label{label4.2}
4\pi\,\frac{dB}{d\Omega_{\vec{n}_p}}(\Lambda^+_c \to p K^- \pi^+
\pi^0/\Lambda^+_c \to pK^-\pi^+)_{\rm
(L)} = 0.87\,(1 + 0.09\,\cos\vartheta).
\end{eqnarray}
Since the coefficient in front of $\cos\vartheta$ is rather small, so
that the angular distribution of the probability of the decays is
practically isotropic. Therefore, one can conclude that in the
four--body mode $\Lambda^+_c \to p + K^- + \pi^+ + \pi^0$ the charmed
baryon $\Lambda^+_c$ seems to be practically unpolarized.

\section{Conclusion}

We have considered the four--body mode of the weak non--leptonic decay
of the charmed $\Lambda^+_c$ baryon: $\Lambda^+_c \to p + K^+ + \pi^+
+ \pi^0$. Experimentally this is the most favourable mode among the
four--body modes of the $\Lambda^+_c$ decays. From theoretical point
of view this mode is rather difficult for the calculation, since
baryonic and mesonic degrees of freedom are not fully
factorized. However, as has been shown in Ref.[2] this problem has
been overcome for the three--body mode $\Lambda^+_c \to p + K^- +
\pi^+$ within the effective quark model with chiral $U(3)\times U(3)$
symmetry incorporating Heavy Quark Effective Theory (HQET) and the
ENJL model [2].

Following [2] we have calculated in the chiral limit the probability
and angular distribution of the probability of the mode $\Lambda^+_c
\to p + K^+ + \pi^+ + \pi^0$ in the rest frame of the $\Lambda^+_c$
baryon and relative to the momentum of the daughter proton. The
probability of the mode $\Lambda^+_c \to p + K^+ + \pi^+ + \pi^0$ is
obtained with respect to the probability of the mode $\Lambda^+_c \to
p + K^+ + \pi^+$. The theoretical prediction $B(\Lambda^+_c \to p K^-
\pi^+ \pi^0/\Lambda^+_c \to pK^-\pi^+)=0.87$ fits well the
experimental data $B(\Lambda^+_c \to p K^- \pi^+ \pi^0/\Lambda^+_c \to
pK^-\pi^+)_{\exp}= (0.68\pm 0.27)$. We would like to accentuate that
in our approach the probability $B(\Lambda^+_c \to p K^- \pi^+
\pi^0/\Lambda^+_c \to pK^-\pi^+)$ does not contain free
parameters. Hence, such an agreement with experimental data testifies
a correct description of low--energy dynamics of strong interactions
in our approach.

The theoretical angular distribution of the probability of the decay
mode $\Lambda^+_c \to p + K^- + \pi^+ + \pi^0$ predicts a rather weak
polarization of the charmed baryon $\Lambda^+_c$. This means that for
the experimental analysis of the polarization properties of the
$\Lambda^+_c$ produced in reactions of photo and hadroproduction the
three--body decay mode $\Lambda^+_c \to p + K^- + \pi^+$ seems to be
preferable with respect to the four--body $\Lambda^+_c \to p + K^- +
\pi^+ + \pi^0$. Nevertheless, the theoretical analysis of polarization
properties of the charmed baryon $\Lambda^+_c$ in the weak
non--leptonic four--body modes like (1) $\Lambda^+_c \to \Lambda +
\pi^+ + \pi^+ + \pi^-$, (2) $\Lambda^+_c \to \Sigma^0 + \pi^+ + \pi^+
+ \pi^-$ and (3) $\Lambda^+_c \to p + \bar{K}^0 + \pi^+ + \pi^-$ with
branching ratios [13]
\begin{eqnarray*}
B(\Lambda^+_c \to \Lambda \pi^+ \pi^+
\pi^-)_{\exp} &=& (3.3\pm 1.0)\,\%,\nonumber\\ B(\Lambda^+_c \to
\Sigma^0 \pi^+ \pi^+ \pi^-)_{\exp} &=& (1.1\pm 0.4)\,\%,\nonumber\\
B(\Lambda^+_c \to p \bar{K}^0 \pi^+ \pi^-)_{\exp} &=& (2.6\pm 0.7)\,\%
\end{eqnarray*}
comeasurable with the branching ratio of the mode $\Lambda^+_c \to p +
K^- + \pi^+ + \pi^0$ is rather actual and would be carried out in our
forthcoming publications.

\section*{Acknowledgement}

The work is supported in part by the Scientific and
Technical Programme of Ministry of Education of Russian Federation for
Fundamental Researches in Universities of Russia.


\newpage

\begin{thebibliography}{9}
\bibitem{[1]} J. D. Bjorken, Phys. Rev. D {\bf 40}, 1513 (1989) and
references therein.
\bibitem{[2]}
Ya. A. Berdnikov, A. N. Ivanov, V. F. Kosmach, and N. I. Troitskaya,
Phys. Rev. C {\bf 60},  015201 (1999).
\bibitem{[3]} E. Eichten and F. L. Feinberg, Phys. Rev. D {\bf 23},
2724 (1981); E. Eichten , Nucl. Phys. B {\bf 4}, (Proc. Suppl.), 70
(1988); M. B. Voloshin, and M. A. Shifman, Sov. J. Nucl. Phys. {\bf
45}, 292 (1987); H. D. Politzer and M. Wise, Phys. Lett. B {\bf 206},
681 (1988); Phys. Lett. B {\bf 208}, 504 (1988); H. Georgi,
Phys. Lett. B {\bf 240}, 447 (1990).
\bibitem{[4]}
M. Neubert,
Phys. Rep. {\bf 245}, 259 (1994);
M. Neubert,
{\it Heavy Quark Effective Theory} CERN--TH/96--292, 
hep--ph/9610385 17 October 1996, Invited talk 
presented at the 20th Johns Hopkins Workshop on 
Current Problems in Particle Theory, Heidelberg, 
Germany, 27--29 June 1996. 
\bibitem{[5]} 
A. N. Ivanov, M. Nagy, and N. I. Troitskaya,
Intern. J. Mod. Phys. A {\bf 7}, 7305 (1992); A. N. Ivanov ,
Phys. Lett. B {\bf 275}, 450 (1992); Intern. J. Mod. Phys. A {\bf 8},
853 (1993); A. N. Ivanov, N. I. Troitskaya, and M. Nagy,
Intern. J. Mod. Phys. A {\bf 8}, 2027, 3425 (1993) ; Phys. Lett. 
B {\bf 308}, 111 (1993) ; Phys. Lett. 
B {\bf 326},  312 (1994); Nuovo Cim. A {\bf 107}, 1375 (1994);
A. N. Ivanov and N. I. Troitskaya, Nuovo Cimento A {\bf 108}, 555 (1995).
\bibitem{[6]}
A. N. Ivanov and N. I. Troitskaya, 
Phys. Lett. B {\bf 342}, 323 (1995); Phys. Lett. B {\bf 345}, 175 (1995); 
A. N. Ivanov  and N. I. Troitskaya,
Nuovo Cim. A {\bf 110}, 65  (1997); A. N. Ivanov, N. I. Troitskaya, 
and M. Nagy, Phys. Lett. B {\bf 339}, 167 (1994); 
F. Hussain, A. N. Ivanov and N. I. Troitskaya, 
Phys. Lett. B {\bf 329}, 98 (1994); Phys. Lett. B {\bf 348}, 609 (1995);
 Phys. Lett. B {\bf 369}, 351 (1996); 
A. N. Ivanov and  N. I. Troitskaya, 
Phys. Lett. B {\bf 390}, 341 (1997); Phys. Lett. B {\bf 394}, 195 (1997);
Phys. Lett. B {\bf 387}, 386 (1996); Phys. Lett. B {\bf 388}, 869 (1996) 
(Erratum).
\bibitem{[7]}
A. N. Ivanov, M. Nagy, and N. I. Troitskaya,
Phys. Rev. C {\bf 59}, 541 (1999).
\bibitem{[8]} T. Hakioglu and M. D. Scadron, Phys. Rev. D {\bf 42},
941 (1990); Phys. Rev. D {\bf 43}, 2439 (1991); R. Karlsen and
M. D. Scadron, Mod. Phys. Lett. A {\bf 6}, 543 (1991); M. D. Scadron,
A {\bf 7}, 669 (1992); Phys. At. Nucl. {\bf 56}, 1595 (1993);
R. Delbourgo and M. D. Scadron, Mod. Phys. Lett. A {\bf 10}, 251
(1995); L. R. Baboukhadia, V. Elias and M. D. Scadron, J. of Phys. G {\bf 23},
1065 (1997); R. Delbourgo and M. D. Scadron, Int. J. Mod. Phys. A {\bf 13}, 657 (1998); A. Bramon, Riazuddin, and M. D. Scadron,
J. of Phys. G {\bf 24}, 1 (1998);M. D. Scadron, Phys. Rev. 
D {\bf 57}, 5307 (1998); L. R. Baboukhadia and M. D. Scadron, Eur. Phys. J. C
{\bf 8}, 527 (1999).
\bibitem{[9]}
A. N. Ivanov, N. I. Troitskaya, M. Faber, M. Schaler 
and M. Nagy, Nuovo Cim. A {\bf 107}, 1667 (1994);
A. N. Ivanov, N. I. Troitskaya and M. Faber, 
Nuovo Cim. A {\bf 108}, 613 (1995).
\bibitem{[10]}
M. Gell--Mann,
Phys. Rev. Lett. {\bf 8}, 214 (1964).
\bibitem{[11]}
B. L. Ioffe,
Nucl. Phys. B {\bf 188}, 317 (1981); Nucl. Phys. B {\bf 191},  591E (1981);
P. Pascual and R. Tarrach,
Barcelona preprint UBFT--FP--5--82, 1982;
L. J. Reinders, H. R. Rubinstein and S. Yazaki,
Phys. Lett. B {\bf 120}, 209 (1983).
\bibitem{[12]} 
M. D. Scadron, 
in {\it ADVANCED QUANTUM THEORY and its
Applications Through Feynman Diagrams}, 
Springer--Verlag, New York, 1st
Edition 1979 and 2nd Edition 1991.
\bibitem{[13]}
D. E. Groom {\it et al.},
Eur. Phys. J. C {\bf 15}, 1 (2000).
\bibitem{[14]}
B. W. Lee and M. K. Gaillard,
Phys. Rev. Lett. {\bf 33}, 108 (1974);
G. Altarelli, G. Curci, G. Martinelli and S. Petrarca,
Nucl. Phys. B {\bf 187}, 461 (1981);
A. Buras,  J.- M. G$\grave{{\rm e}}$rard and R. Ruckl,
Nucl. Phys. B {\bf 268}, 16 (1986);
M. Bauez, B. Stech and M. Wizbel,
Z. Phys. C {\bf 34}, 103 (1987).
\bibitem{[15]} 
M. D. Scadron and L. R. Thebaud, Phys. Phys. Rev. D
{\bf 8}, 2190 (1973); R. E. Karlsen and M. D. Scadron, Phys. Rev. D
{\bf 43}, 1739 (1991); M. D. Scadron and D. Tadi${\acute{c}}$, {\it
Hyperon Nonleptonic Weak Decays Revisited}, hep--ph/0011328 November 2000,
to appear in J. of Phys. G.
\bibitem{[16]} 
V. Elias and M. D. Scadron, Phys. Rev. D {\bf 30}, 647
(1984); Phys. Rev. Lett. {\bf 53}, 1129 (1984).
\end{thebibliography}
\end{document}